# A Search for Optical Laser Emission from Proxima Centauri

G. W. Marcy[1*]

[1] *Center for Space Laser Awareness, 3388 Petaluma Hill Rd, Santa Rosa, CA, 95404, USA*
*www.spacelaserawareness.org*

ABSTRACT

A search for laser light from Proxima Centauri was performed, including 107 high-resolution, optical spectra obtained between 2004 and 2019 with the HARPS spectrometer. Among them, 57 spectra contain multiple, confined spectral "combs", each comb consisting of 10 closely-spaced frequencies of light. The spectral combs, as entities, are themselves equally spaced with a frequency separation of $5.8 \times 10^{12}$ Hz, rendering them unambiguously technological in origin. However, the combs do not originate at Proxima Centauri. The comb light is Doppler shifted such that the Earth's orbital motion is removed, as if the comb source is at Earth. Also, the spectra of several other stars show similar combs at similar light frequencies. Detailed examination of the raw images shows the combs come from optical ghosts of an interferometric etalon filter at the telescope, about which there is no record in the data logs nor any mention in published papers about these spectra. Thus, the 107 spectra of Proxima Centauri show no evidence of technological signals, including 29 observations between March and July 2019 when the candidate technological radio signal was captured by Breakthrough Listen. This search would have revealed lasers pointed toward Earth having a power of 20 to 120 kilowatts and located within the 1.3au field of view centered on Proxima Centauri, assuming a benchmark laser launcher having a 10-meter aperture. Smaller lasers would also have been detected, but would require more power.

## 1   INTRODUCTION

Speculative models of the Milky Way Galaxy suppose that it may contain a network of mutually communicating spacecraft probes stationed near stars (e.g. Bracewell 1973; Freitas 1980; Maccone 2014; Maccone 2018; Hippke 2021; Gertz 2021). The probes may communicate by lasers to optimize privacy, achieve high bandwidth, and minimize payload, and the laser wavelengths may be ultraviolet, optical, or infrared (Schwartz and Townes 1961, Zuckerman 1985, Hippke 2018).

One signature of interstellar laser communication is the narrow range of wavelengths, nearly monochromatic, of its light (e.g., Naderi et al. 2016, Su et al. 2014, Wang et al. 2020). Searches for such laser communication have been carried out using high-resolution spectra of over 5000 normal stars of spectral type F, G, K, and M, yielding no detections and no viable candidates (Reines & Marcy 2002; Tellis & Marcy 2017). A similar search for laser emission from more massive stars of spectral type O, B, and A, has also revealed no viable candidates (Tellis et al. 2021, in preparation). These laser searches involved examining high-resolution spectra, $\lambda/\Delta\lambda > 60000$, in the wavelength range $\lambda = 3600$ to $9500$ Å, for monochromatic emission lines. The required laser power for detection is 1 to 10 MW, assuming a diffraction-limited laser emitter consisting of a benchmark 10-meter aperture. None was found.

Meanwhile, searches continue for radio-wave signatures of other technologies in the Milky Way, currently pursued by the Breakthrough Prize Foundation (e.g., Lebofsky et al. 2019, Price et al. 2020), the UCLA SETI Group (Margot et al. 2021), and by other radio telescopes. Using the Parkes radio telescope pointed at Proxima Centauri, the Breakthrough Prize Foundation LISTEN team discovered a signal at 982 MHz consistent with a technological origin (Sheikh et al. 2020, 2021; O'Callaghan & Billings 2020, Overbye 2020, Loeb 2021, Koren 2021).

This radio signal and the proximity of its source of only 4.24 light years promote Proxima Centauri to the top of the target list in the sky for SETI observations. A technology there, only slightly advanced compared to ours, could detect specific human actions on Earth, digest them, and send an electromagnetic response to Earth within 8.5 years. This paper describes a search for laser emission in 107 optical spectra of Proxima Centauri. A following paper describes another search, with a novel telescope, for laser emission coming from the Solar gravitational lens point of Proxima Centauri.

*E-mail: geoff@spacelaserawareness.org

## 2  OBSERVATIONAL METHOD

I retrieved all 107 spectra of Proxima Centauri obtained with the HARPS spectrometer between 2004 Feb 25 and 2019 July 13 and kindly made available on the ESO public data archive (archive.eso.org). HARPS is a fiber-fed, high-resolution echelle spectrometer at the ESO 3.6 m telescope at the La Silla Observatory in Chile (Mayor et al. 2003). I obtained the archived 1D extracted spectra, along with a wavelength scale calibrated in the solar system barycentric frame (European Southern Observatory, HARPS User Manual, 2011). These wavelengths are Doppler shifted from the observatory frame to account for the orbital motion of the Earth around the Sun, the motion around the Earth-Moon barycenter, and the rotation of the Earth, accurate to within 0.1 ms$^{-1}$. The spectra have a nominal resolution of $\lambda/\Delta\lambda = 115000$ and span wavelengths, $\lambda = 3781 - 6913$ Å, and are rebinned to 0.01 Å per pixel.

The spectra of Proxima Centauri were taken between 2004 and 2019 within four different observing programs, all listed and described in the Acknowledgments section. All spectra were obtained on the 3.6-meter ESO telescope, with exposure times of typically 15 minutes and reduced by the standard HARPS pipeline, yielding signal-to-noise ratios in the stellar light-flux continuum of typically 30 per resolution element.

Figure 1 shows a typical spectrum of Proxima Centauri on a log intensity scale to exhibit the flux from blue to red in the stellar continuum. The spectrum contains thousands of atomic and molecular absorption lines (appearing as upside-down "grass"), typical for stars of spectral type dM5.5. The spectrum also contains many emission lines that may compromise the search for laser emission. I also examined 95 HARPS spectra of "comparison" stars that are located within 2 degrees of Proxima Centauri (Vmag=11.1), namely HD125881 (Vmag=7.25, G2V), HD126793 (Vmag=8.2, G0). Both comparison stars were observed with similar exposure times as was Proxima Centauri, and both stars are sufficiently faint to reveal night-sky lines or artifacts that could masquerade as laser emission.

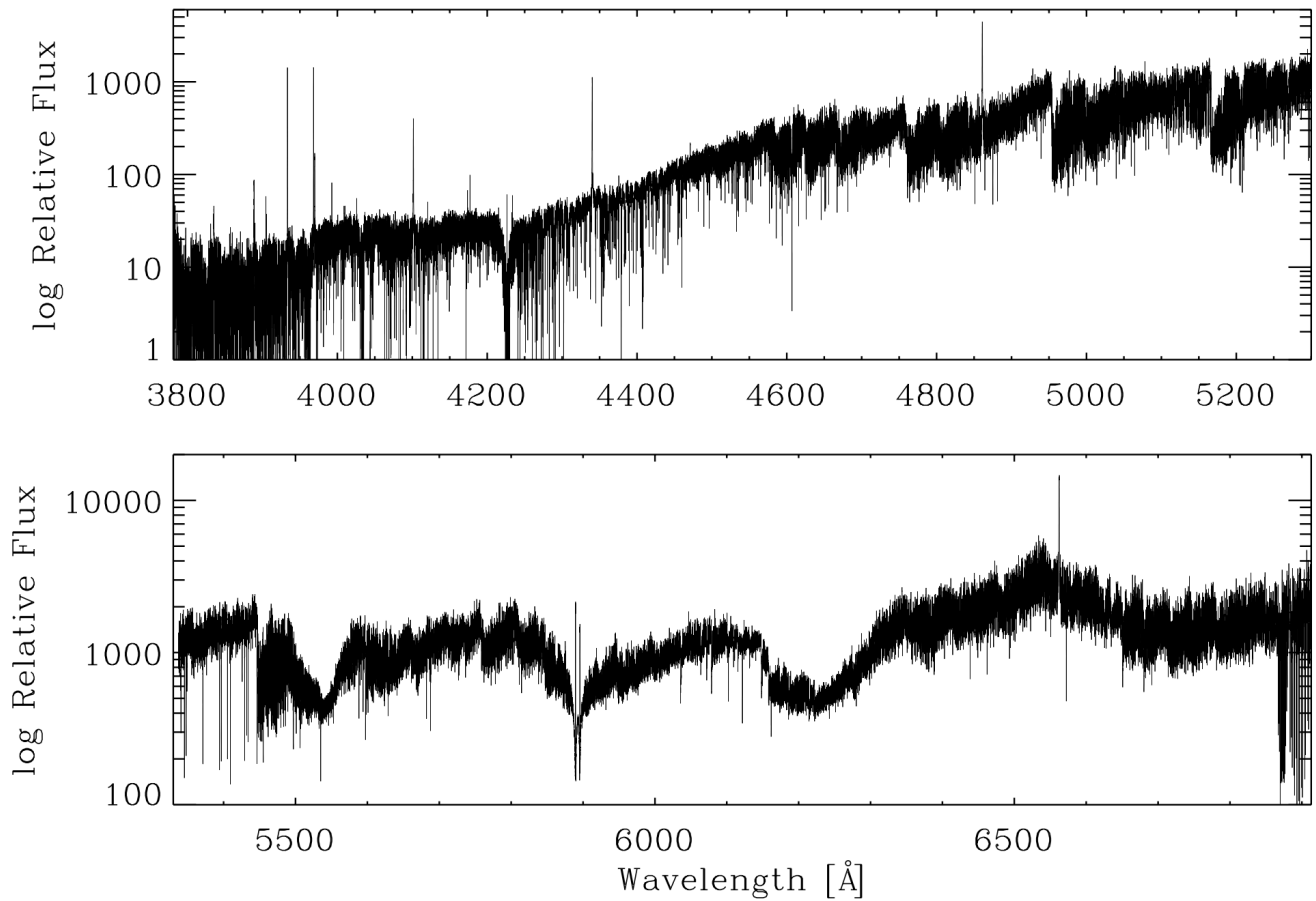

*Figure 1. Representative optical spectrum of Proxima Centauri with no flare occurring, plotted as log photons per pixel vs. wavelength. Thousands of atomic and molecular absorption lines are visible, typical of spectral type dM5.5. The emission lines deserve examination and interpretation.. The signal-to-noise ratio varies from 5 to 40 (blue to red) due to the spectral energy distribution of Proxima Centauri and throughput of the spectrometer.*

I verified the spectral resolution by examining spectral lines that are intrinsically narrower than the nominal instrumental profile, notably the two night-sky emission lines [OI] 5577 and [OI] 6300 and telluric $O_2$ absorption lines. Those diagnostic lines displayed a FWHM of 6 pixels, corresponding to 0.06 Å, consistent with a spectral resolution of at least $\lambda/\Delta\lambda$ =100,000, as shown in Figure 2.

These diagnostic spectral lines normally exhibit widths slightly greater than the instrumental profile due to small pressure broadening and thermal broadening occurring in the Earth's atmosphere along the line of sight to the star. Both effects cause broadening of less than 20% over that of the instrumental profile, which explains their widths being slightly greater than implied by the nominal resolution of $\lambda/\Delta\lambda$ = 115,000. *Any laser emission line at a wavelength, $\lambda$, must have a spectral width, $\Delta\lambda$, at least as broad as given by $\lambda/\Delta\lambda$ = 115,000, as verified by these diagnostic spectral lines in Figure 2.*

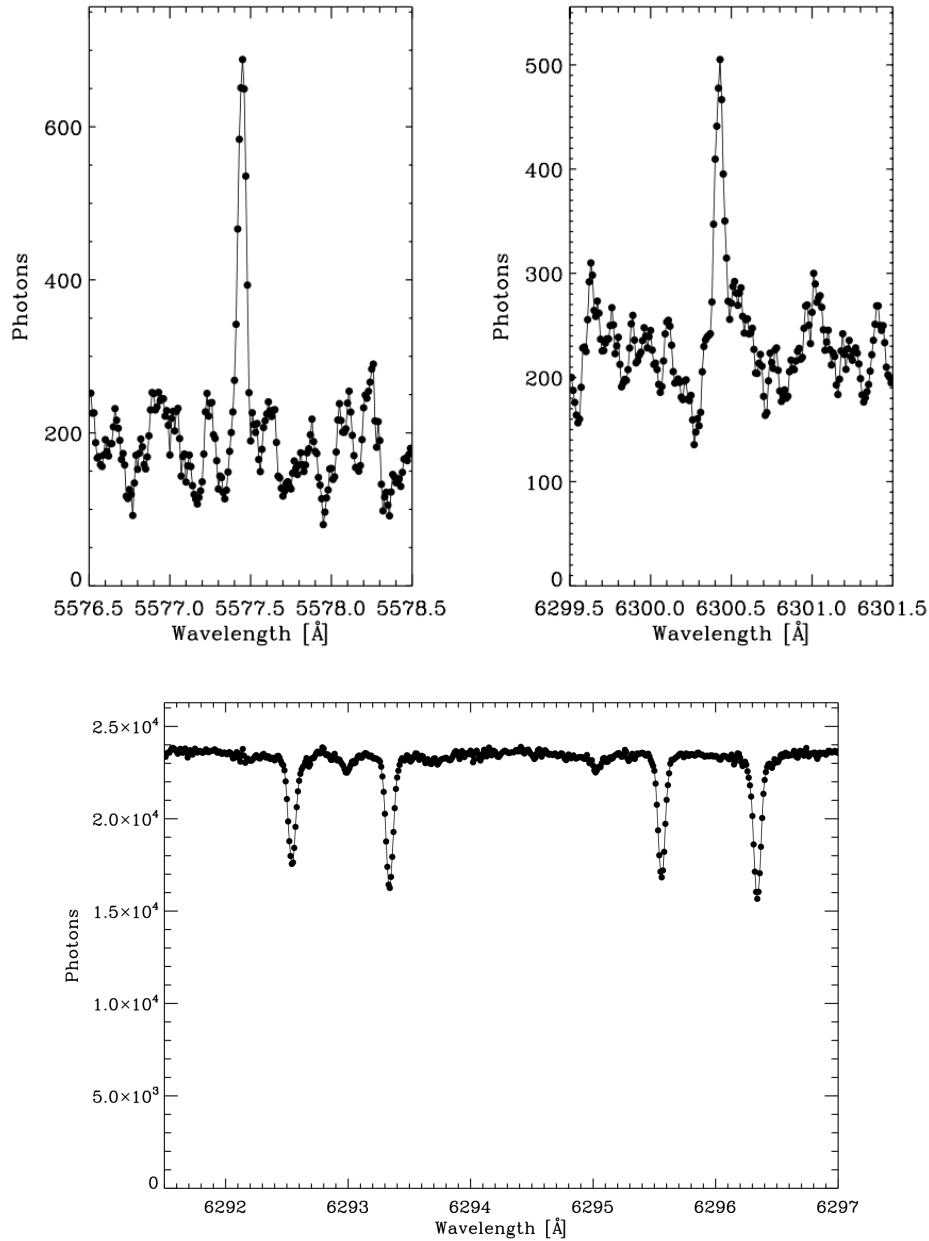

*Figure 2. Proxies of the instrumental profile of HARPS.* **Top:** *Observed night sky emission lines, [OI] 5577 Å and [OI] 6300 Å in the spectrum of Proxima Centauri.* **Bottom:** *Absorption lines due to molecular oxygen in the Earth's atmosphere. All proxies exhibit FWHM = ~6 pixels (0.060 Å), implying resolution better than $\lambda/\Delta\lambda$ = 100000. Any laser emission line must have FWHM at least 6 pixels wide.*

The entrance fiber of HARPS has a diameter that projects to 1.00 arcseconds on the sky. The original circular fibers were replaced with octagonal cross-section fibers in May 2015 of similar size cross-section. Proxima Centauri is 1.302 pc from Earth (Gaia DR3, 2020). Thus, the fiber collects light from a cone extending outward with a 1 arcsecond opening angle covering a footprint of diameter 1.30 au at Proxima Centauri. Any light emitted from within that cone and directed toward Earth could be detected in these spectra. Light originating outside that cone obviously cannot be detected. Thus, this search for laser emission from Proxima Centauri can identify laser sources located within a cone-shaped volume having a diameter of 1.3 au there and extends into the background behind the star, along the line to Earth. The small fiber aperture prevents us from assessing, in the raw CCD images, the angular extent on the sky of any spectral emission, as enabled by a long entrance slit (Tellis & Marcy 2017).

## 3    SPECTRAL EMISSION DURING FLARES ON PROXIMA CENTAURI

I searched for laser emission lines in all 107 spectra of Proxima Centauri. The emission could arrive as a single frequency of light emitted by a single laser. Alternatively, the emission could arrive as a series of closely spaced frequencies of light that are smeared together by the instrumental profile of HARPS, creating an apparent spectral emission line that is broader than the instrumental profile. We thus search for non-astrophysical emission lines that are as broad or broader than the instrumental profile (Figure 2).

I first identified the astrophysical emission lines associated with flares and chromosphere on Proxima Centauri. All of the spectra show prominent Balmer lines in emission, typical of heating by magnetic fields in the upper atmospheres of M dwarfs (e.g., Reiners & Basri 2008, Hawley et al. 2014). The Balmer emission lines are 10x more intense during times of flaring compared with times of no flares, as shown in Figure 3, in the bottom panel compared to the top panel. During flares, emission also occurs at the NaD lines 5892Å, HeI 5876Å, HeII 4868Å, and the CaII H&K lines, all typical during flares of M dwarfs.

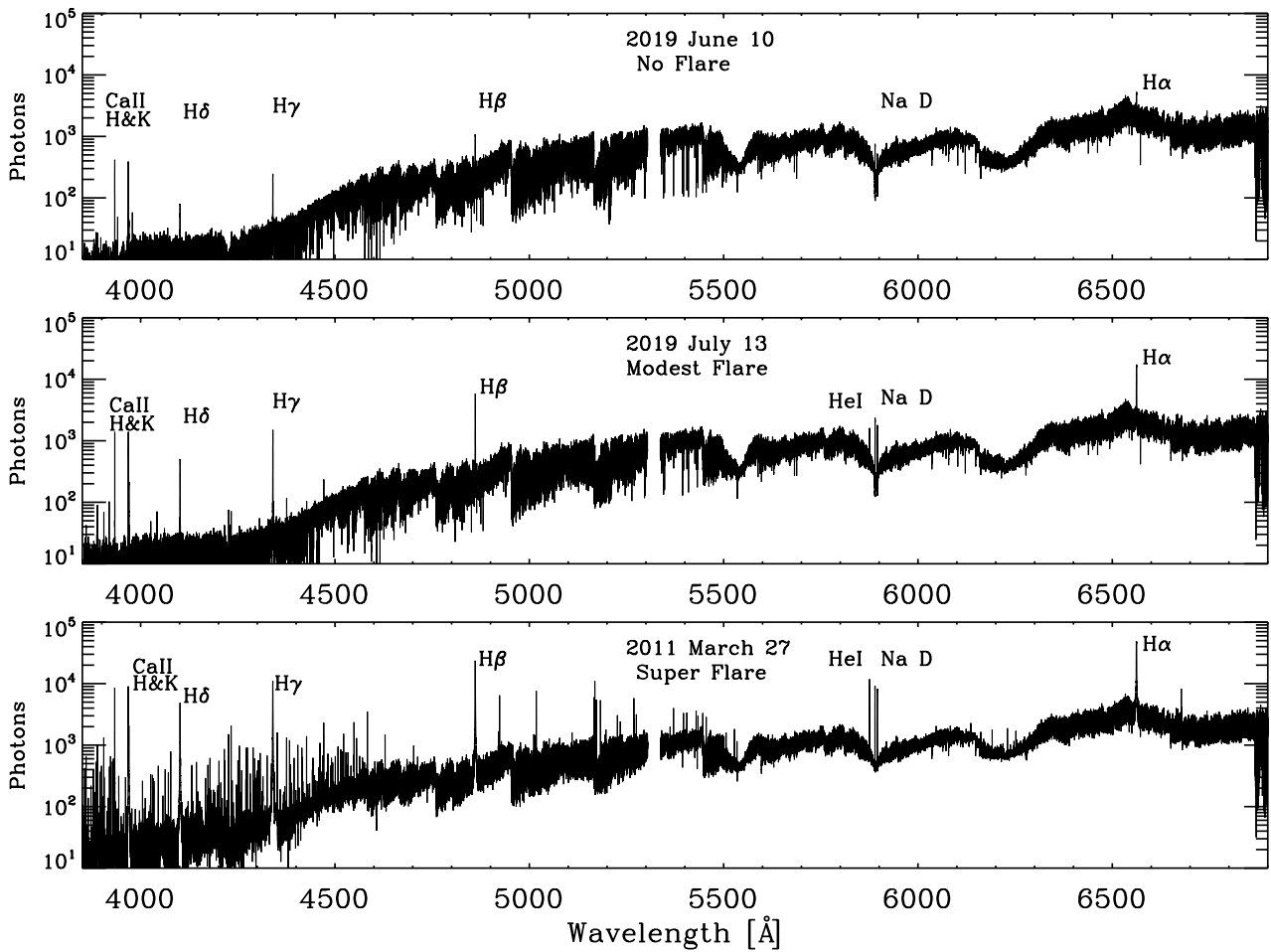

*Figure 3. Three representative spectra of Proxima Centauri, during times of no flare (top), modest flare (middle) and super flare (bottom). The vertical axis is photons per 0.01 Å, on a log intensity scale. Emission in Balmer lines, NaD, HeI, and CaII H&K are prominent, and many other transitions from heavy elements appear during super flares.*

Figure 4 shows the detailed shape of the spectral emission lines from HeI and HeII during one of the flares of Proxima Centauri. Atmospheric modeling of optical emission lines from Proxima Centauri indicate gas in excess of 10,000K in its chromosphere, its flare regions, and in an extended corona-like hot envelope (Pavlenko et al. 2017). The flares on Proxima Centauri occur many times per day, last for many minutes, and emit bursts of electromagnetic radiation at radio to X-ray wavelengths, including brightening at visible wavelengths by a factor of 100 during its "super flares" (e.g. Howard et al. 2018; Zic et al. 2020). I searched for high excitation and high ionization emission in the HARPS echelle spectra of Proxima Centauri, and indeed there are several prominent examples. Figure 4 shows emission from ionized and neutral helium, requiring temperatures over 10000K and excitation of 20 eV, consistent with UV and X-ray from flares, and radio-wave emission from cyclotron processes.

Remarkably, hundreds of narrow emission lines also appear in spectra of Proxima Centauri during the strongest flares, as shown in the bottom panel of Figure 3 and in the zoom of wavelength regions shown in Figures 5 and 6. Figures 5 and 6 show emission lines arising from common atoms such as iron and titanium normally formed in absorption in the photosphere due to a declining source function. These remarkable lines in emission include prominently FeI, FeII, TiII (e.g. Davenport et al 2012). The observed widths of these lines are typically slightly more than 6 pixels, corresponding to Doppler broadening of ~3 kms$^{-1}$ (thermal and bulk fluid motion) convolved with the instrumental profile (Figure 2).

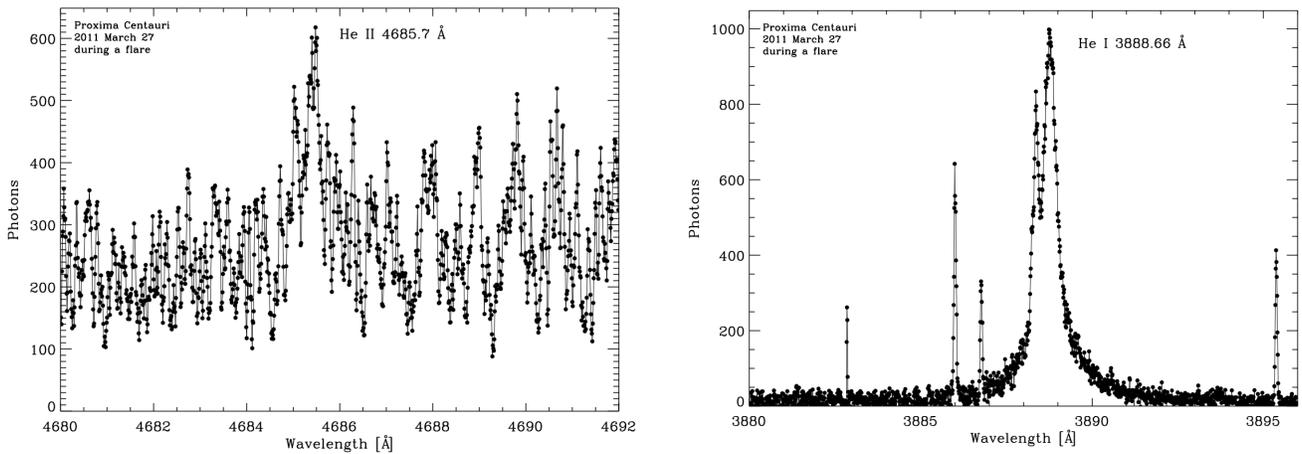

Figure 4. Emission from He II 4685.7 and He I 3888.66 during strong flares that occur in 5 out of 95 spectra of Proxima Centauri, indicating plasma temperatures of ~10000K and production of UV and X-ray photons, consistent with direct measurements (e.g., Howard et al. 2018 and Zic et al. 2020). The narrow emission lines are from heavy elements (Fig 5).

These spectral emission lines shown in Figures 5 and 6 arise from heavy atoms in the magnetically heated upper atmosphere where the radiative transfer source function is inverted. These lines appear commonly in emission in other M dwarf flare stars and T Tauri stars, and they appear in the spectra of Solar-type stars in absorption. These narrow emission lines from heavy elements in Proxima Centauri have roughly the same shape as spectral lines arising from technological laser emission. *Thus, the identification of candidate laser emission from Proxima Centauri is complicated by the hundreds of emissions lines, having widths only slightly greater than the instrumental profile, that occur naturally and commonly during flares of Proxima Centauri.*

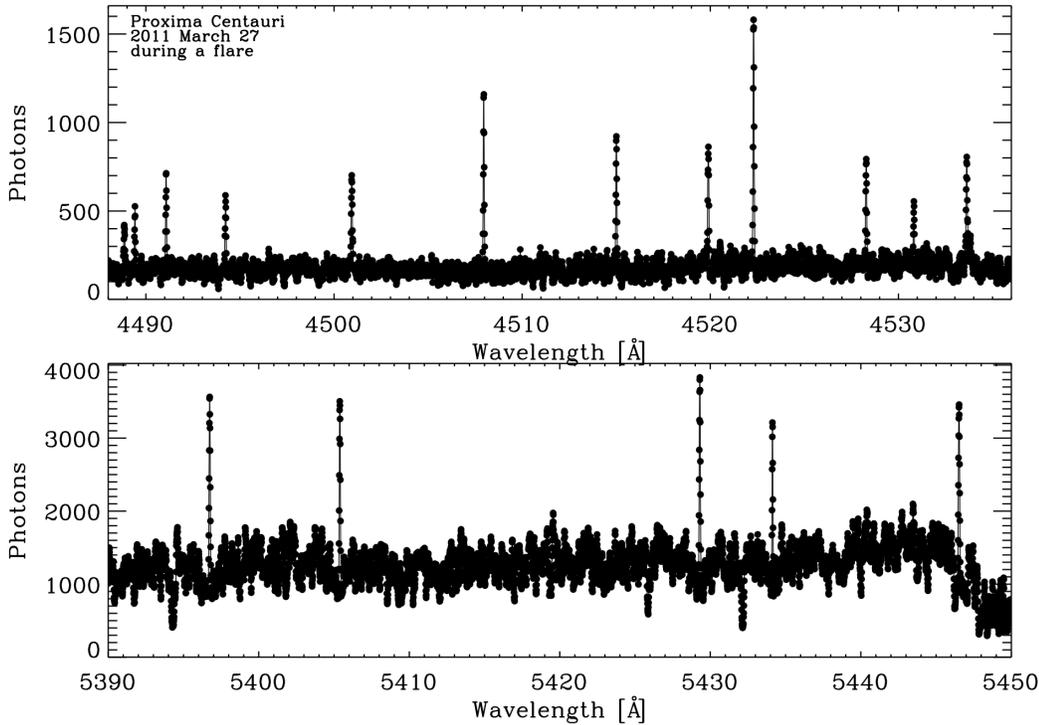

*Figure 5. Emission lines from atomic transitions that normally appear in absorption from the photosphere, arising from FeI, FeII, and TiII during a flare on Proxima Centauri. The lines have widths (FWHM) of ~6 pixels consistent with the instrumental profile of the HARPS spectrometer convolved with small thermal broadening in gas at a temperature of 10000K. These natural emission lines have widths indistinguishable from those of technological lasers, making optical SETI difficult.*

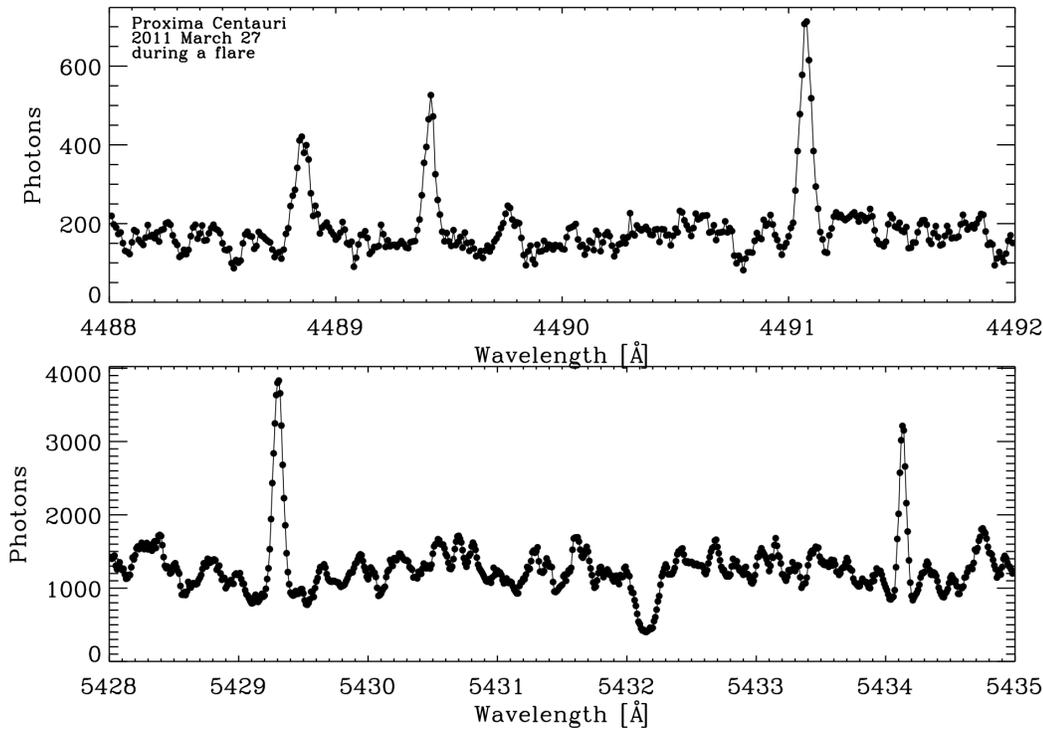

*Figure 6. Magnified view of a few emission lines appearing in Figure 5. The lines arise from Fe and Ti in the 10,000K flare regions. The widths of the lines are ~6 pixels, caused by the instrumental profile convolved with modest thermal broadening at flare temperatures over 10000K, making these lines difficult to distinguish from the technological, monochromatic emission expected from an optical laser. Lasers of kW and MW power may have intrinsic line widths up to many Angstroms due to structural and thermal imperfection caused by high energy density.*

## 4  SEARCH FOR LASER LINES

I established a method to search for laser emission in the 107 high-resolution spectra of Proxima Centauri. The algorithm must avoid the rich molecular bands in absorption and dozens of spectral lines in emission that occur sporadically and often in the magnetic chromosphere and flare regions. This spectral and temporal variability forces adoption of a generous threshold for candidate laser lines at 5 times the stellar continuum intensity. I created a continuum-normalized spectrum by computing a median-smoothed spectrum with a smoothing width of 1000 pixels corresponding to 10 Å. I divided each spectrum of Proxima Centauri by its median-smoothed spectrum to produce a spectrum with a pseudo-continuum normalized to 1.0.

Setting the detection threshold for laser lines is not simple. Qualitatively, emission lines (from chromosphere or flares) can be securely detected against the diverse molecular features if they extend to a normalized intensity of 5.0 and higher, suggesting a threshold for laser emission. This threshold is not rigorously optimized by a statistical criterion, nor can it be. Thousands of absorption lines from molecules in the photosphere prevent the identification of a definite spectral "continuum". Also, there are hundreds of emission lines that come and go with flare activity. Further, direct cosmic ray hits often rise as high as 10x the stellar counts in the pixels. The spectral landscape does not offer a statistically definable topography against which to set a statistically rigorous threshold for culling laser-line candidates. Eyeball examination of the 107 spectra of Proxima Centauri reveals that a threshold of 5x the pseudo-continuum will yield a few dozen candidate laser lines.

Specifically, the detection criterion for emission lines is as follows. The observed peak intensity within a 0.01 Å pixel must be 5x the median intensity of the star's spectrum within 0.01 Å, with the running median computed in 10 Å segments. The star's 10 Å median intensity is, of course, a function of wavelength. Further, there is a criterion for the line width in wavelength. Candidate emission lines must have a width (FWHM) between that of the instrumental profile (~0.05 Å) and 5 Å. The upper limit in line width at 5 Å is set by the median smoothing width noted above. I note that broad emission, up to 5 Å wide, can occur either by a tight cluster of narrow laser lines smeared by the spectrometer's instrumental profile or by one fat laser line.

I searched the full wavelength extent of all 107 spectra of Proxima Centauri for any "emission line" meeting the detection threshold defined above. Roughly 1/3 of the spectra have such emission. Many of these emission lines were narrower than the FWHM of 6 pixels that characterizes the instrumental profile (see Figure 2). These narrow "emission" features are likely due to elementary particles hitting the CCD directly, including cosmic rays (often secondary muons) and beta and gamma rays coming from natural radioactivity in the Earth and observatory structures. Figure 7 exhibits two of these events from elementary particles. They are too narrow to be from light that passed through the spectrometer, eliminating them as candidate laser lines.

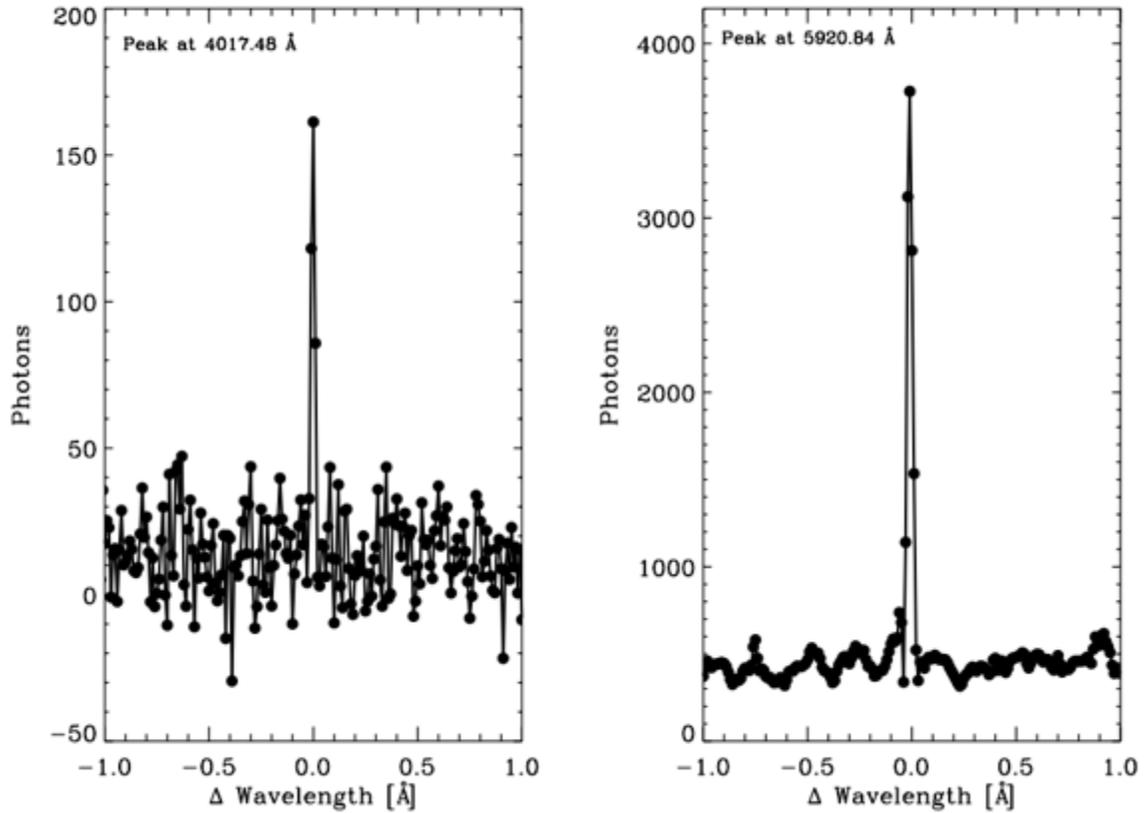

*Figure 7. Two representative emission features more intense than our threshold for laser emission from Proxima Centauri. However, they have widths narrower than the ~5-pixel FWHM of the instrumental profile of the HARPS spectrometer. Such features did not pass through the spectrometer and are likely due to high energy particles hitting the CCD directly. We reject such features from further consideration.*

I searched all 107 spectra of Proxima Centauri for emission lines having a peak intensity at least 5x the continuum intensity and having widths with FWHM > 5 pixels, consistent with the instrumental profile. Six candidates emerged, all shown in Figure 8. Among them, close examination revealed five having widths slightly less than the minimum, FWHM>6 pixels, required to be consistent with the instrumental profile of the HARPS spectrometer (Figure 2). In addition, some candidate lines have wings or double peaks inconsistent with the instrumental profile, making them not credible candidates of laser emission that entered the telescope and optical system. They are likely to be due to direct impacts on the sensor by elementary particles.

The remaining candidate consists of a cluster of approximately 10 emission lines, all within 1.5 Å in the wavelength interval, 4184.0 Å – 4185.5 Å. The "comb" of lines exhibits an arch-like envelope of intensity, weakest at the short and long wavelengths and stronger near the central wavelengths. This comb appears in all 57 spectra taken during 10 nights between 2013 May 4 and May 14 (UT). The comb is accompanied by regions of excess, noise-like light at UV and blue wavelengths both shortward of 4000 Å and less so longward. In contrast, none of the 49 spectra of Proxima Centauri taken on other dates between 2004 and 2019 exhibits that emission comb.

This emission comb is not mentioned in the published paper by Anglada-Escudé et al. (2016) that presents these spectra, nor does that paper mention using any unusual calibration lamp or light source. I searched the literature for atoms or molecules with transitions within this wavelength interval of the comb, but did not find any. Nor are there stellar absorption lines at those comb wavelengths, either in Proxima Centauri or in G and K stars, except for a few CN and CH transitions.

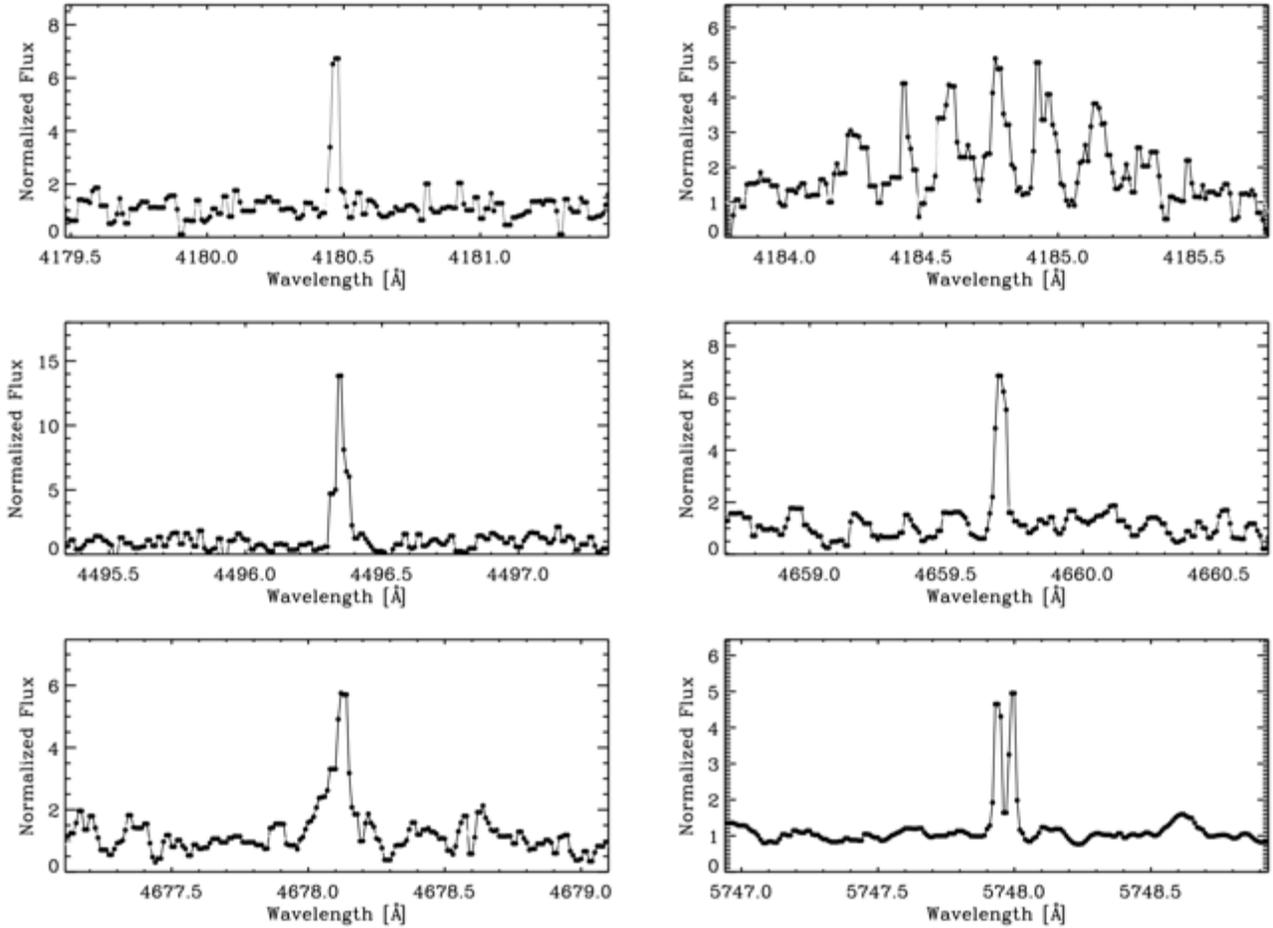

*Figure 8. Unvetted candidate laser lines that emerged from our search of 107 high-resolution spectra of Proxima Centauri. All spectra are normalized to place the stellar continuum at 1. Five candidates have widths less than that of the instrumental profile FWHM = 6 pixels (see Fig. 2), thereby eliminating them as light that passed through the optics. They may be elementary particles that hit the CCD. The last candidate (upper right) consists of a comb of perhaps 10 emission lines between 4184 – 4185.5 Å that were unidentified. This emission comb appears in 57 of 107 spectra of Proxima Centauri, but not in the 75 spectra of comparison stars, HD125881 or HD126793.*

Figure 9 shows all 57 occurrences of this emission comb in the spectrum of Proxima Centauri. The emission occurs with different intensities, varying by a factor of ~3, from exposure to exposure. Figure 10 displays the 57 comb emission spectra overplotted and summed. These show that the wavelengths of the teeth of the comb were nearly (but not exactly) the same, within 0.05 Å, in all exposures on 10 nights. The occurrence of the comb is not correlated with the intensity of emission lines that come from the chromosphere and flares in Proxima Centauri, such as H-$\alpha$, H-$\beta$, CaII H&K, HeI 4686 Å, HeI 3888 Å, CaI 4277 Å, nor the neutral metal lines in the blue. Even during super flares, the emission comb at 4185 Å is often absent, and conversely, when the emission comb is present, the flare diagnostics are often absent.

The wavelength 4185 Å is not used in laser guide star adaptive optics, for which the sodium D line transition at 5895 is used in most astronomical applications. The Apache Point laser operates at 532nm (harmonic of Nd:YAG), and Robo-AO uses a laser at 355 nm (Jensen-Clem et al. 2017). The Four-Laser Guide Star System uses a seed laser with a stable output at 11780 Å and subsequent frequency doubling to yield a wavelength of 5890 Å (Calia, Hackenberg, Holzlöner, et al., 2014). The 3[rd] harmonic is at 3926 Å, not consistent with 4185 Å. Other laser guide star systems use nonlinear sum-frequency generation of natural transitions of Nd:YAG crystals at 1064 nm and 1319 nm, the harmonics of neither of which reside at 4185 Å. Also common are Nd-doped fiber lasers at 938 nm and commercial 1583 nm fiber laser that also do not have frequency doubling or tripling at 4185 Å. It remains possible that some experimental laser guide star system that employs sum-frequency generation was being used, either at the La Silla Observatory, a nearby facility, or a satellite.

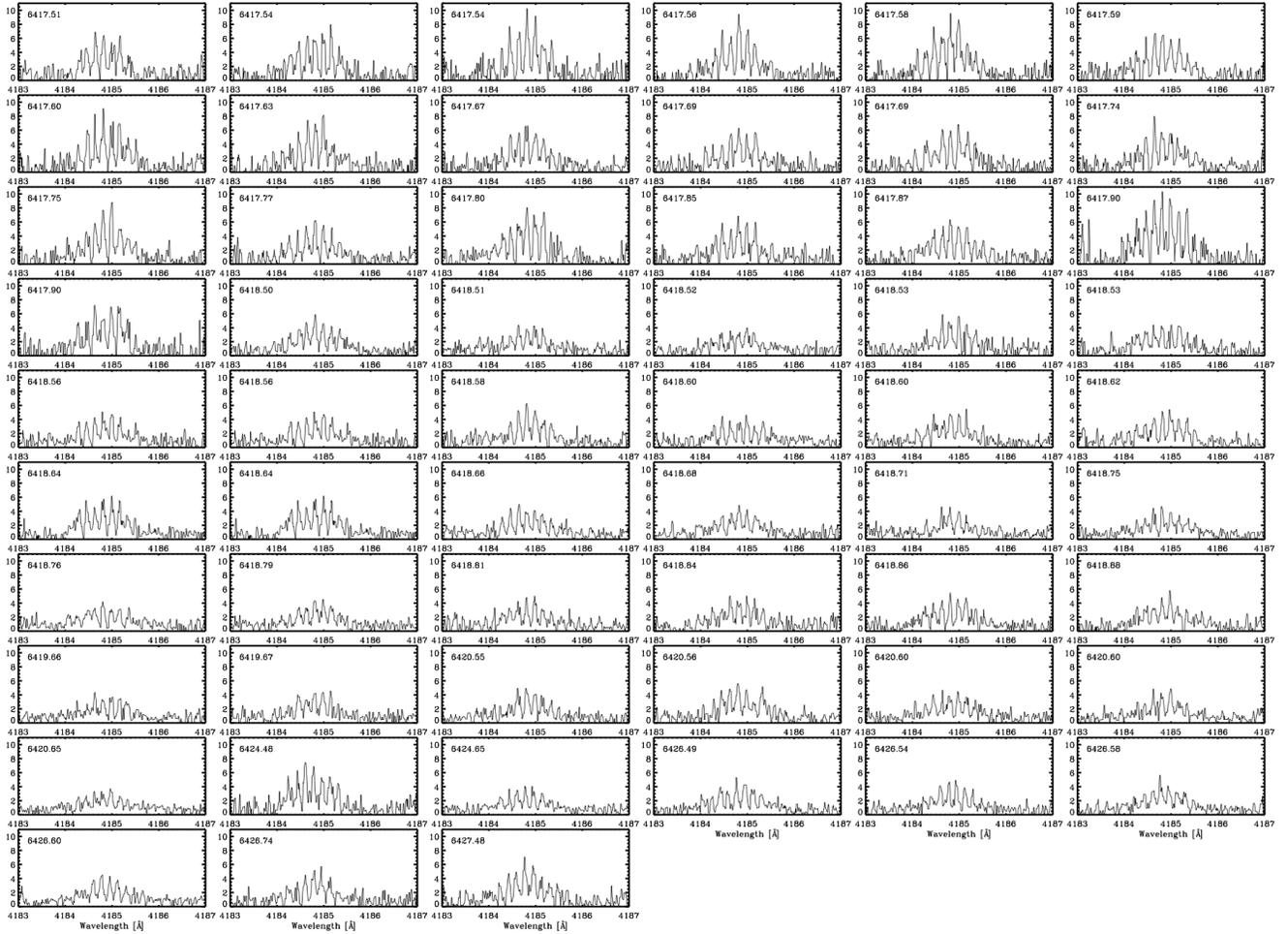

*Figure 9. All 57 HARPS spectra taken between 4 to 14 May 2013 of Proxima Centauri, centered on wavelength 4184.8 Å and in chronological order. At upper left is JD-2450000. The spectra are scaled to put the stellar continuum at 1. All spectra of Proxima Centauri taken during these 10 days exhibit the comb of emission.*

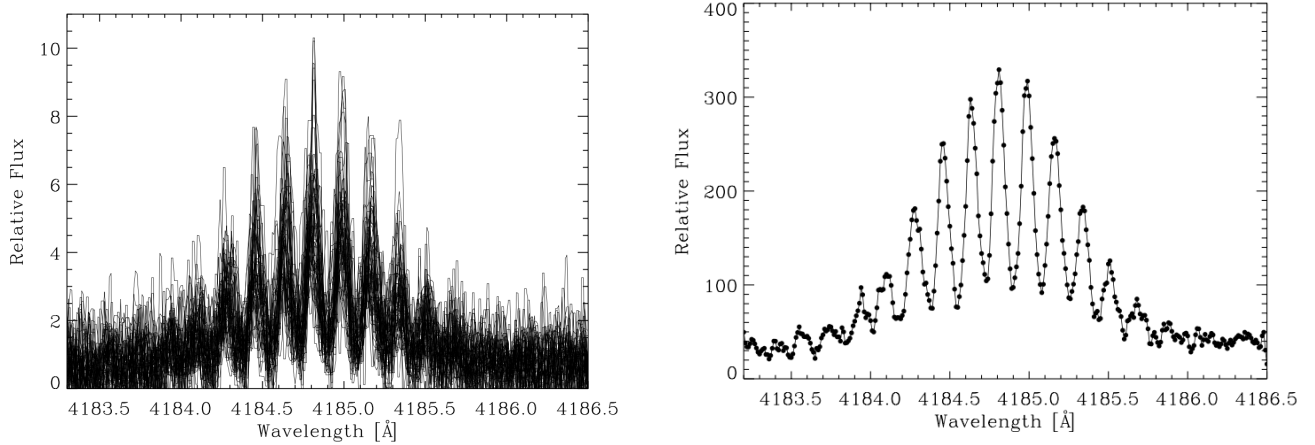

*Figure 10. **Left:** Overplot of all 57 spectra of Proxima Centauri near 4184.8 Å, of Proxima Centauri taken between 5 – 14 May 2013. The comb fringes appear at nearly the same wavelengths on all 10 nights, within ~0.05 Å, implying a Doppler drift <4 km s$^{-1}$. **Right:** The sum of the 57 spectra at the comb region. The fringes are nearly equally spaced by 0.175 Å and have shapes only slightly wider than the instrumental profile (Figure 2), indicating narrower intrinsic shapes.*

The sum-frequency pairs, defined by $1/\lambda = 1/\lambda_1 + 1/\lambda_2$, permit combinations of commonly used laser wavelengths to convert to a different frequency. For example, 1064nm mixed with 1535nm yields 628 nm. I considered all commonly used laser

wavelengths, namely, 532nm, 589nm, 938nm, 1064nm, 1178nm, 1300nm, 1319nm, 1550nm, 1583nm. None of the pairs yields a resulting wavelength near 418.5 nm as observed.

The "comb-like" emission appears in all 57 spectra of Proxima Centauri, spanning 10 nights of that observing run. Its declination, -62° 40', implies the ESO 3.6-m telescope was always pointed within 28 degrees of the south celestial pole. Any source of the 4185 Å comb in the southern direction could, in principle, be the origin. Figure 11 shows that the ESO 3.6-m telescope is located at the far southern end of the ridge, with no buildings or telescopes south of it. No facility is located south of the 3.6-m where light could have been directed toward the atmosphere above the 3.6-m dome allowing light to scatter into the 3.6-m. It remains possible that light comb comes from a launch site located north of the ESO 3.6-m telescope. We have found no record of such a facility, but experiments with laser guide star adaptive optics may have occurred. Still, the persistence of the comb within all 57 observations of Proxima Centauri makes it less likely that an outside facility could explain all of the combs.

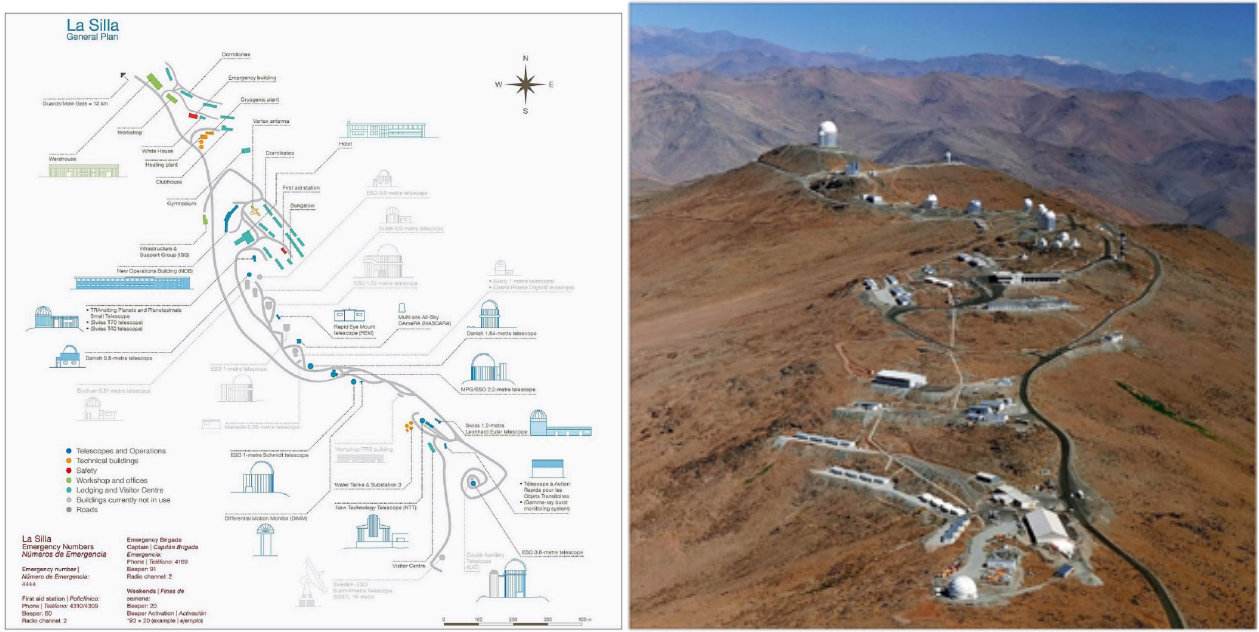

*Figure 11. Left: Map of the La Silla Observatory, with north up. The 3.6-m telescope is at the bottom-right, at the south end of the ridge. Right: a photo of the La Silla ridge, looking south. The ESO 3.6-metre telescope is at the top. The 3.6-m telescope was pointing within 29 deg of due south and within 60 deg of the horizon, where no telescopes are located, nor any facilities in the distance. Map and photo are courtesy ESO and C. Madsen.*

The strong magnetic field of strength 500 g raises the possibility of Zeeman splitting of some atom (Reiners & Basri 2008) at Proxima Centauri. Also possible is some rovibronic transition of a molecule. A satellite-based optical laser system at 4185 Å is possible, but any satellite between altitudes of low earth orbit and geosynchronous orbit seems unlikely because such a satellite could not hover on the line-of-sight to Proxima Centauri so far from the celestial equator.

I measured the Doppler drift (relative to the first night) of the spectral comb in all 57 spectra between 4 – 14 May 2013 using a simple cross-correlation algorithm, accurate to 0.01 Å (=0.7 kms$^{-1}$). The HARPS pipeline wavelength scale establishes the wavelength of each tooth of the comb. These wavelengths are different in each observation, and the change in wavelength can be interpreted as a non-relativistic Doppler shift.

Figure 12 (top panel) shows the resulting Doppler drift of the comb during the 57 observations of Proxima Centauri. The Doppler shift of the comb increases by 3.5 kms$^{-1}$ during the 10 days. The teeth have assigned wavelengths that become smaller during 10 days, plotted here as a positive quantity. Figure 12 (middle panel) shows the Doppler effect caused by the Earth's orbital motion around the Sun. The plot shows the component of the Earth's velocity away from Proxima Centauri at the time of each observation, shown as individual dots. Figure 12 (middle panel) shows that the Earth's velocity toward Proxima Centauri changes by 3.5 kms$^{-1}$, similar to the change in the Doppler shift of the comb shown in the top panel. The bottom panel displays an overplot of the top two panels. The Doppler drift of the comb and the Doppler effect due to the Earth's orbital motion are consistent with each other.

The wavelengths at each pixel for each exposure are those as if the observatory were both located at the centre of mass of the Solar System and acquiring light coming from Proxima Centauri. The original observatory-frame wavelengths (from wavelengths of emission lines of a thorium-argon lamp) were changed to those as if the observations of Proxima Centauri were performed at the Solar System barycenter.

Figure 12 (bottom panel) shows the similarity of the 3.5 kms$^{-1}$ Doppler shift, during 10 days, of the comb and the Earth's orbital velocity toward Proxima Centauri. In the Solar System barycenter frame, the comb light changes its radial velocity toward Proxima Centauri the same as the 3.6-meter telescope, *consistent with the source being located at the observatory.*

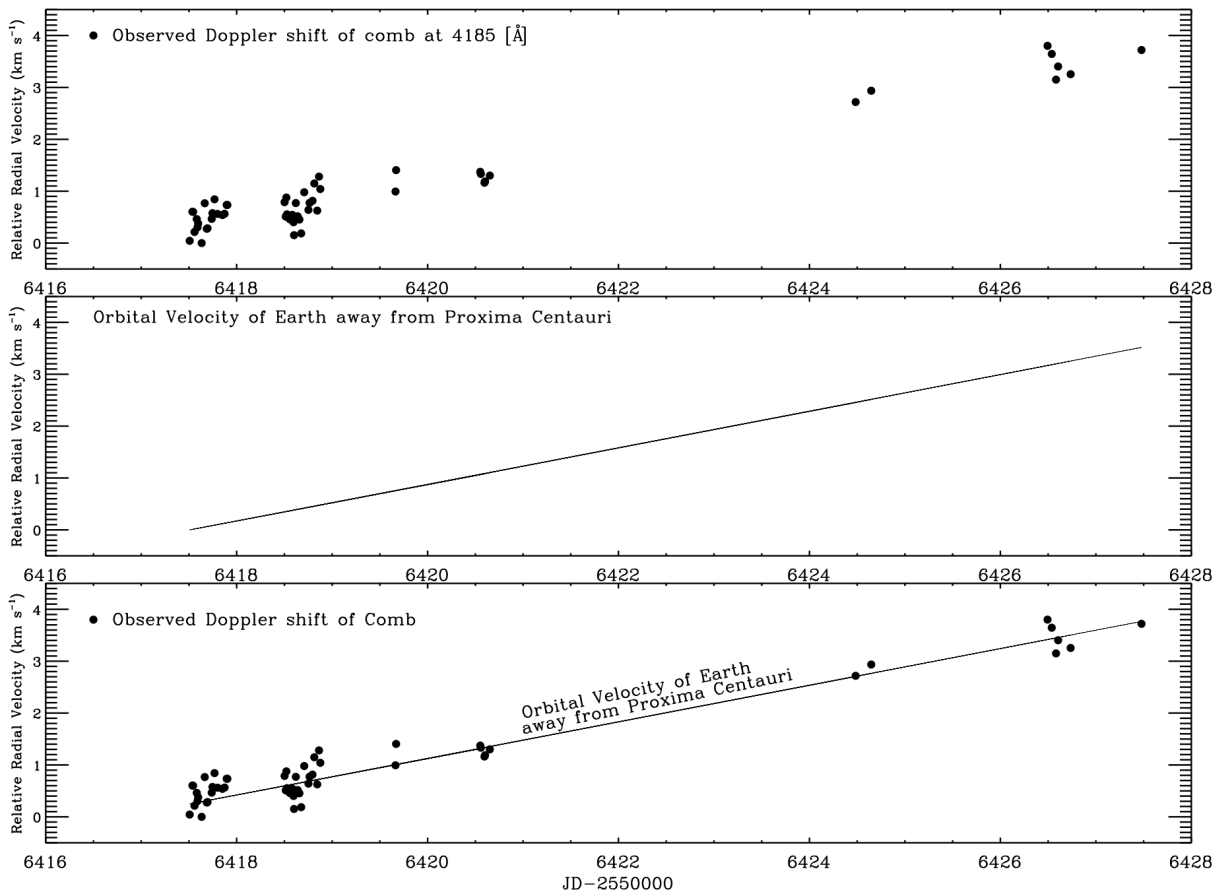

*Figure 12.* ***Top:*** *Measured Doppler shift of the unidentified spectral comb at 4185 Å.* ***Middle:*** *The calculated velocity of the 3.6-meter telescope away from Proxima Centauri due to the Earth's motion.* ***Bottom:*** *The measured Doppler shift of the comb overplotted on the orbital velocity of Earth away from Proxima Centauri. The agreement shows the source of the comb light is changing its radial velocity toward Proxima Centauri exactly as is the Earth.*

The spectra of Proxima Centauri taken 4 – 14 May 2013 contain other spectral comb-like emission features besides the one at 4185 Å. I co-added the 57 spectra of Proxima Cen during those 10 days to build up signal-to-noise, enabling a search for other combs. I securely identified 11 other combs. Another three or four combs appear in the spectrum to the eye, but their shapes are so distorted that these combs may not exist. We display all 12 combs in Figure 13. All combs have a similar spacing of the comb teeth, and all exhibit 8 to 10 teeth having intensities that rise and fall within the envelope ~1.5 Å long. There appears to be no "envelope" of the intensities of the combs, as entities, from UV through the blue region of the spectrum.

As entities, the combs appear to occur at regular wavelength intervals of ~32 Å, which is apparently more regular in delta frequency of 5.8x10$^{12}$ Hz, as judged from the wavelength of the peak tooth in the comb (Table 1). A more accurate fiducial wavelength position of the combs could be constructed by a matched filter approach. We notice distorted combs at wavelengths of the predicted locations, assuming a constant comb spacing of 5.8x10$^{12}$ Hz, but their distorted shapes render them uncertain. The wavelengths and frequencies of the centres of the 12 combs in the spectra of Proxima Centauri are given in Table 1 and shown in Figure 13.

TABLE 1.  Wavelengths and Frequencies of 12 Combs in Proxima Centauri

| $\lambda$ (Å) | 3992.70 | 4025.99 | 4058.48 | 4090.93 | 4121.53 | 4184.81 | 4221.53 | 4257.70 | 4293.30 | 4361.33 | 4402.33 | 4442.35 |
|---|---|---|---|---|---|---|---|---|---|---|---|---|
| $\nu \times 10^{14}$ Hz | 7.5086 | 7.4464 | 7.3867 | 7.3280 | 7.2712 | 7.1638 | 7.1015 | 7.0406 | 6.9832 | 6.8736 | 6.8096 | 6.6488 |

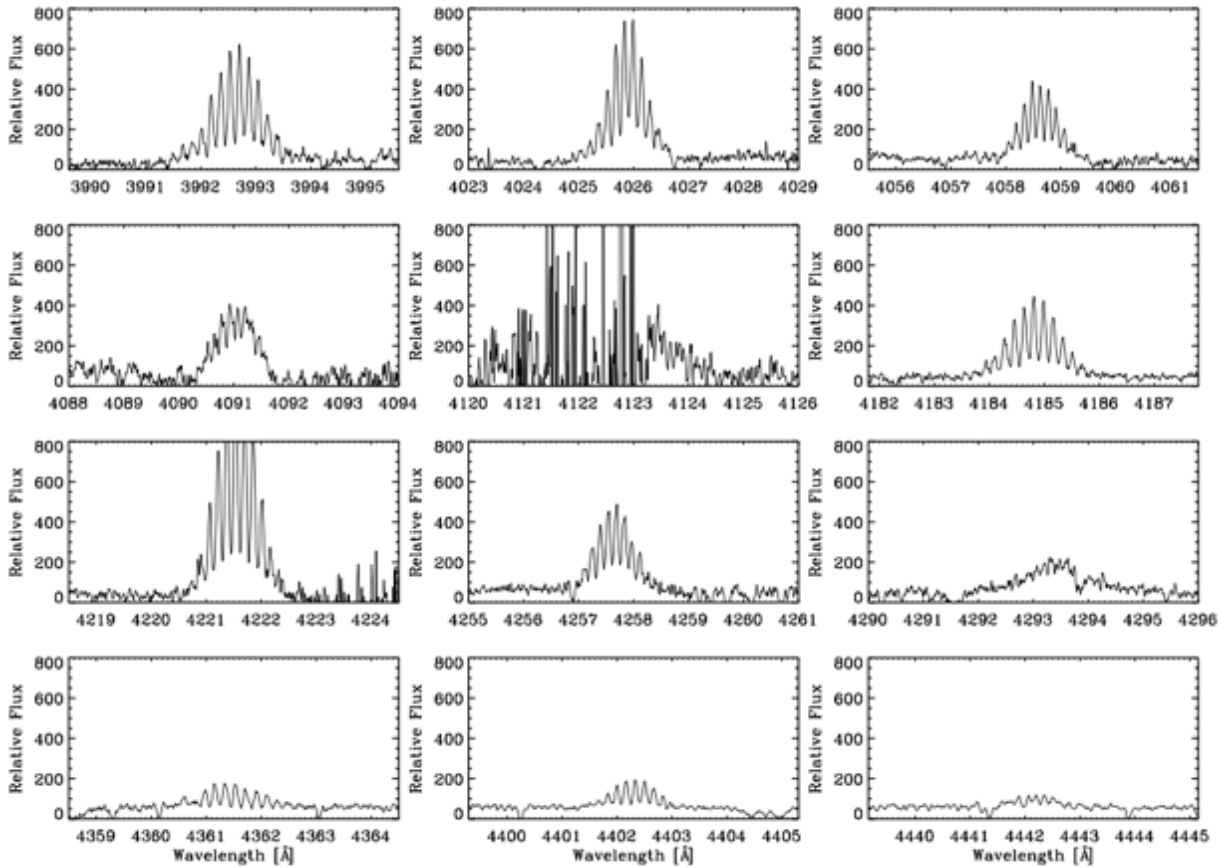

*Figure 13. A compilation of 12 different emission combs discovered in the spectra of Proxima Centauri during 4 - 14 May 2013. The combs as entities are separated from each other in wavelength by ~32 Å and a nearly constant frequency interval of $5.8 \times 10^{12}$ Hz. Some combs are missing from the equal-separation pattern, and some shortward of 3992 Å have considerable noise, as do some longward such as 4123 Å shown here in the centre, 2nd row.*

## 5 EXPLANATIONS OF THE COMBS IN THE SPECTRA OF PROXIMA CENTAURI

Each of the 57 high-resolution spectra of Proxima Centauri obtained during May 2013 contains 12 comb-like spectral emission features distributed throughout near UV and blue wavelengths. The emission combs cannot be of astrophysical origin because their central frequencies (or wavelengths) are more regular than from any naturally occurring atomic or molecular source of light. The regularity of the combs as entities immediately rules out atomic transitions, molecular spectral bands, and Zeeman splitting as the cause of the comb emission. The frequency spacing of the 12 combs is inconsistent with any astrophysical sources. The lack of Doppler drift relative to the Earth implies the comb source is likely some interferometric optical device on Earth, likely at the ESO La Silla observatory and perhaps at the 3.6-meter telescope itself.

One possible explanation is that some telescope instrument was shining a beacon into the sky containing the comb-like emission, allowing that light to scatter off the air into the ESO 3.6-meter telescope. If so, the beacon must have been active during all 57 observations of Proxima Centauri and of other stars. That telescope could be located at the La Silla Observatory or some other facility located off La Silla. If located on the Earth's surface, that facility would not need to adjust the light frequencies of its comb emissions to ensure no Doppler shift change, night after night. It is not plausible that the machine emitting the comb light was located off the surface of the Earth, because all of the observed stars, located many degrees apart in the sky, exhibit the combs.

To ascertain the location of the comb light, I searched the spectra of other M-type dwarf stars similar in brightness to that of Proxima Centauri, with B magnitudes 10 to 13, and observed during May 2013. Figure 14 shows spectra of some M dwarfs observed during the same era with HARPS. Indeed, they all exhibit the combs. The only exception is the brightest star, GJ-191. Its blue brightness, Bmag = 10.4, provides enough stellar photons to dwarf the comb light. Our sampling suggests all of the hundreds of HARPS spectra obtained during 10 days (at least) contain these combs throughout the near-UV and blue portions of the spectrum. As these stars were located all over the sky, a beacon from another telescope could not be the explanation for the combs.

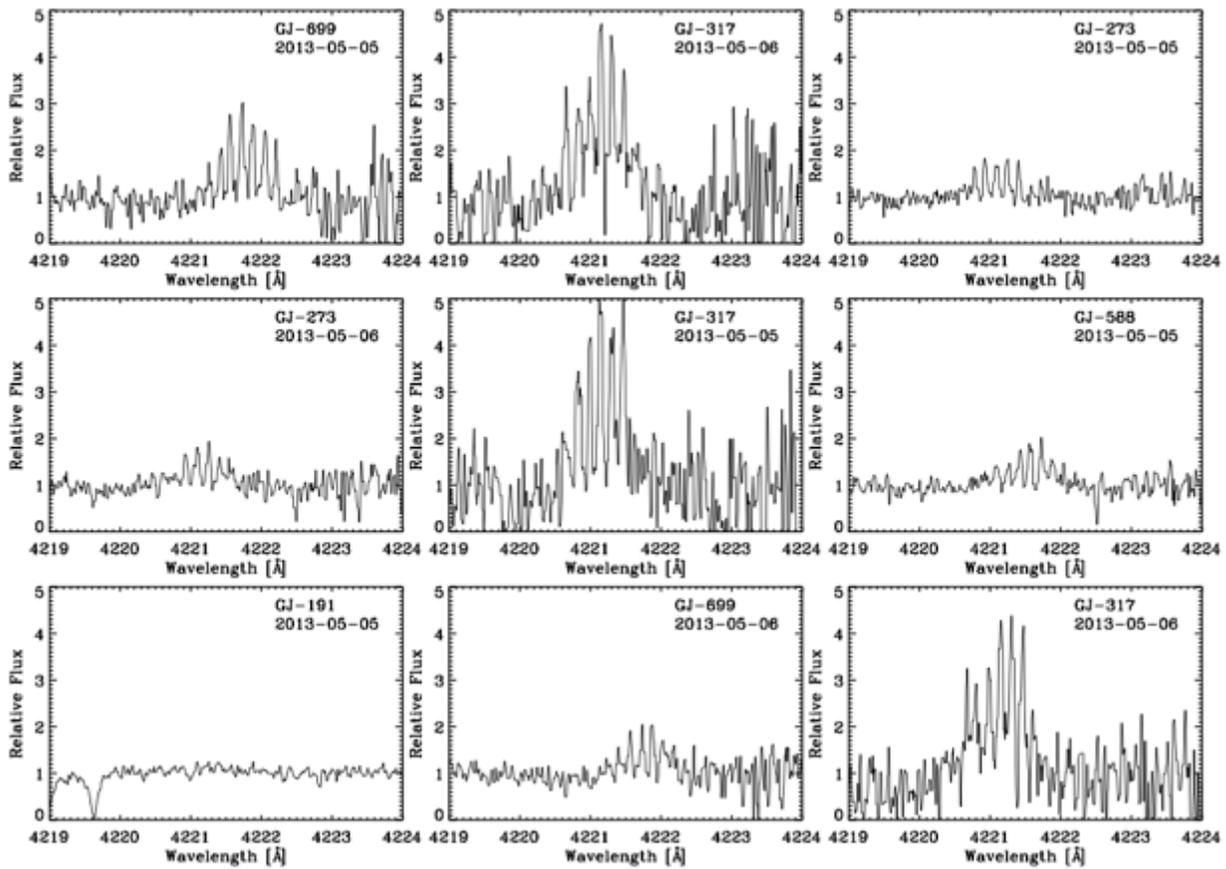

*Figure 14. Spectra of other M-type dwarf stars obtained with the HARPS spectrometer during May 2013, with a zoom at 4222 Å. The "GJ" star number and date of observation are shown at upper right. The combs are clearly visible in other stars, with contrast consistent with the B magnitude. (GJ-191 is the brightest, hence a diluted comb.) Thus, the comb light comes from the observatory, not Proxima Centauri or any particular star.*

Another possibility is that the ESO 3.6-meter telescope itself was experimenting with, or using, Fabry-Perot etalons or laser combs, with an option to use such calibration lamps to obtain high Doppler precision. The FITS headers of the spectra in May 2013 contain no reference to a Fabry-Perot etalon, a simultaneous th-ar lamp, nor any other fringe-producing device such as a laser comb. The paper by Anglada-Escudé et al. (2016) describing these spectra of Proxima Centauri does not contain a mention of a Fabry-Perot etalon nor a laser comb. That paper describes each HARPS spectrum, "the wavelength of which is calibrated using a hollow cathode lamps (Th-Ar)." Anglada-Escudé et al. (2016) describe this observing run, "May 2013: 143 spectra obtained on three consecutive nights between 4 May and 7 May and 25 additional spectra between 7 May and 16 May with exposure times of 900 s." There is no mention of any source of "combs," "fringes," "Fabry-Perot etalons," or any interferometric device.

To investigate the origin of the combs, I communicated with the principal investigator of the observations of Proxima Centauri in May 2013, Guillem Anglada-Escudé in January 2021. He kindly offered much useful information, and he wrote that no interferometric devices were used. He kindly offered the possibility that "ghost spectra come from the secondary fiber." But he stated that he "switched off the calibration light completely … to avoid that precisely." I thank Dr. Anglada-Escudé for his generous information.

Nonetheless, a general possibility is that some optical glass in the HARPS spectrometer behaved as a Fabry-Perot etalon, allowing internal reflections of light to yield interference fringes. To test for such optical origins of the combs, I examined the raw CCD images of Proxima Centauri. One CCD image is shown in Figure 15.

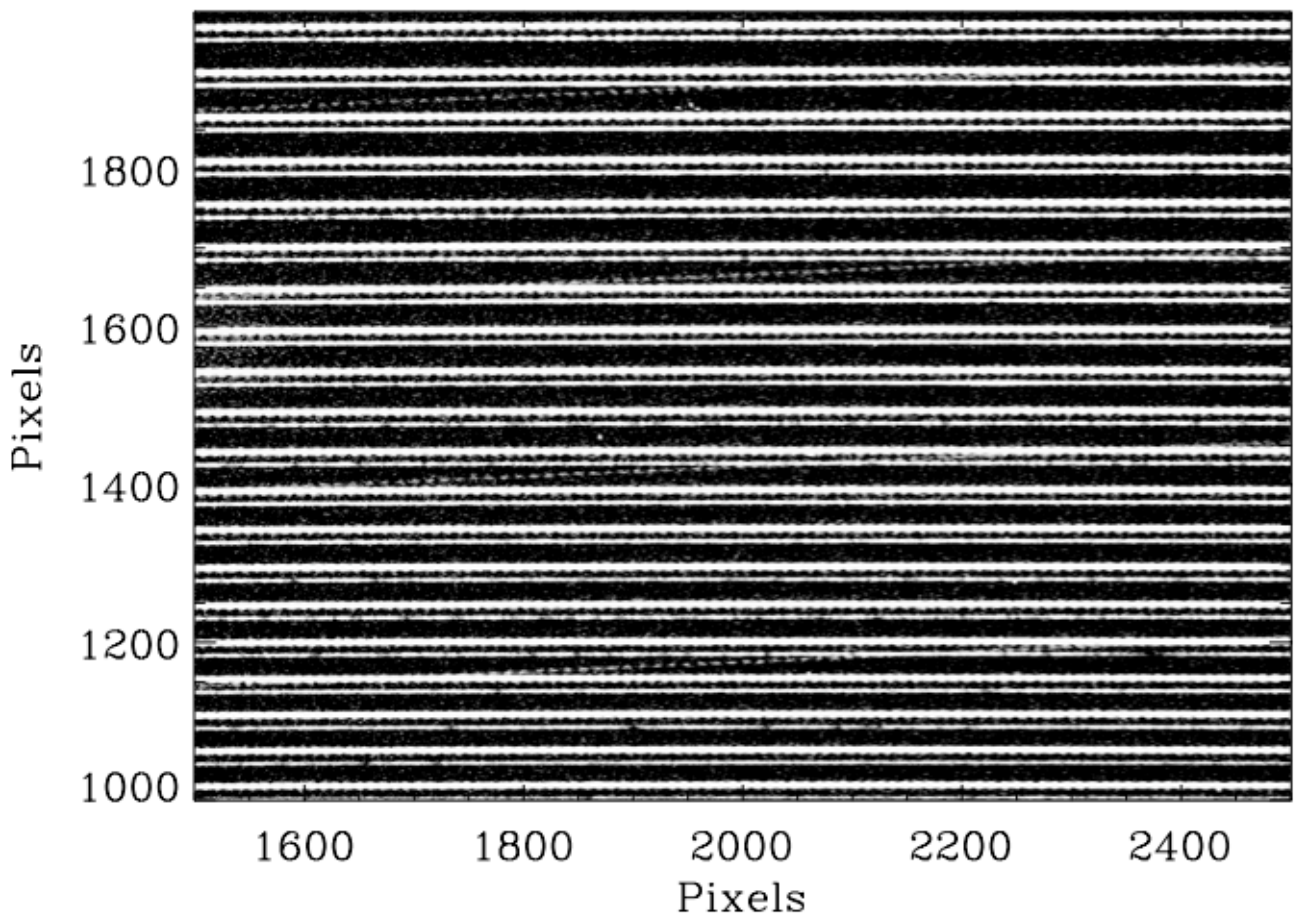

*Figure 15. A portion of the raw CCD image of the HARPS echelle spectrum of Proxima Centauri, with longer wavelengths at the top. The double horizontal stripes are the spectral orders, containing two spectra. Fainter and below is the spectrum of the star, as expected. Above each of them is the spectrum of the Fabry-Perot wavelength calibration lamp, which was not previously known to be present. Faint diagonal stripes are caused by the ghost produced by the unintended internal reflections in the grism cross dispersor. These ghosts send faint interferometric fringes slashing through the spectrum of Proxima Centauri.*

Figure 15 shows the expected horizontal stripes that are the spectral orders of the stellar spectrum from the HARPS spectrometer. Parallel and above each stellar spectral order is light from the Fabry-Perot etalon and its lamp, which was not supposed to be on, according to published papers and the astronomers present during that run. Furthermore, there are faint diagonal stripes of light at the bottom, slicing through the spectral orders (Figure 15). These are optical "ghosts" caused by internal reflections within the grism cross-dispersor of the HARPS spectrometer. Neither the Fabry-Perot fringes nor the ghosts were known to exist in these spectra. The diagonal ghosts consist of interferometric fringes that overlap and contaminate the spectra of Proxima Centauri, constituting the combs.

In summary, the Fabry-Perot etalon and its lamp were left on during the entire 10-night observing run in May 2013, which was not known. In principle, the resulting interferometric fringes should not be a problem. But that light reflected within the cross-disperser grism produces ghosts in the form of diagonal strips of light that pierce the stellar spectrum, causing the combs.

I queried the chief scientist of the HARPS spectrometer, Francesco Pepé. He kindly investigated the images and stated, "I confirm that simultaneous Fabry-Perot was used. I confirm that what you observe is on both 1D and 2D spectra. It is localized to about 1 Å width and occur once per echelle order at 'arbitrary' position." The source of the comb is now confirmed to be the Fabry-Perot etalon and its lamp, along with an optical ghost.

The 12 comb fringes in each of hundreds of spectra obtained during that 10-night run may have slightly compromised the Doppler measurements used to search for planets. The combs appear at wavelengths that are artificially Doppler shifted in each spectrum, on time scales of hours, days, and months, as observed in Figure 12. The comb emission does not participate in the Doppler shift caused by the Earth's motion around the Solar System's barycenter, while the light from Proxima Centauri (and all observed stars) is Doppler shifted by the Earth's motion. The 12 combs cover a total of 20 Angstroms of spectrum and are ~5 times brighter than the light from Proxima Centauri, implying spectral contamination that is Doppler shifting in the solar system barycenter frame. Thus, the 12 combs that are stationary in wavelength may dilute the measured Doppler shift of the spectrum. One amelioration is that the spectrum "template" against which the Doppler shifts are measured presumably does not contain comb-like fringes. If so, the Doppler measurements may suffer from extra noise rather than systematic errors. I suspect the Doppler measurements were affected only slightly.

## 6  THE CHALLENGE OF SETI

This paper describes the lengthy effort to find an explanation for the "comb" feature in Proxima Centauri. As the feature surely comes not from that star but from the HARPS spectrometer at the observatory, the report of this circuitous effort could be removed from this paper. However, the effort represents a common challenge of SETI observations, worthy of discussion.

A challenge of SETI is that every unexpected attribute of the data might be a "signal." Every data quirk must be examined in sufficient depth to explain it in terms of known phenomena. All unexpected results in the data, i.e., "non-canonical phenomena" (Singam et al. 2020), could be caused by such effects as the unexpected performance of novel equipment, an unexpected natural phenomenon here on Earth, human error in data-taking technique or analysis, or none of the above. SETI is the search for "none of the above."

SETI research thus requires that every data quirk be understood, *a posteriori*. One cannot go back in time to the moment the quirk appeared in the original data to confirm, with care, the arrival of a signal. Such time-travel would permit the use of other telescopes, spectrometers, cameras, or calibrations that were specifically designed to confirm or reject the original candidate signal. Such hypothesis-driven experiments are profoundly different, and simpler, than searching for unknown phenomena (Singam et al. 2020). When a data-quirk is detected, any actual electromagnetic wave comprising the signal has already passed the Earth at light speed.

One cannot let any unexpected data quirks go unvetted because the first extraterrestrial signal will be unprecedented. Any unexplained, potential signal must be examined exhaustively, lest the cherished signal remain dubious. Cutting-edge scientific equipment and the experiments at the frontier are particularly fraught with unexplained quirks in the data. SETI observations, per force, examine domains where no one has gone before.

Equally daunting is that the Bayesian "prior" probability of credibility is low for any claim of a detected extraterrestrial signal. A casual poll of astronomers shows that the prior is less than 10%, no matter the apparent quality of the claimed SETI detection, for many reasons.

1. Astronomers have searched the night sky for 40 years, every night, using advanced imaging and spectroscopy, at all wavelengths and employing dozens of professional 3-10 meter class telescopes. The yield is no technological signals. Thus, a large slice of the multi-dimensional search space *observable by modern technology* has already been observed. See Wright, Kanodia, Lubar (2018) for a broader view.

2. Artifacts of other technologies have been sought, including images of the Moon and Mars and the Lagrange points in the Solar System. Those implicit searches yielded no artifact.

3. The past 60 years of explicit SETI searching has yielded no detections (Wright, Kanodia, Lubar 2018).

4. The past century of exploration of the Earth has yielded no non-human eroded spacecraft or structures on land nor under the ocean (Wright 2017, Shostak 2020).

5. Technological entities will likely build optical or IR interferometric telescope arrays in space having kilometer baselines enabling the detection of chlorophyll, oxygen build-up, and city lights on Earth, encouraging them to send probes here (Zuckerman 2002). If so, the lack of detections of those probes may reflect their desire to remain hidden. The prior must reflect this possibility.

Some of these five reasons may be over-stated, but probably not all of them. The product of these priors can't be quantified, but yields a probability of successful detection much less than 1, to be multiplied by the claimed probability of veracity of each reported signal. The final prior probability is much less than unity, and is the origin of the common intuition that SETI may require decades or centuries for success.

# 7    LASER DETECTION THRESHOLDS FOR PROXIMA CENTAURI

This search for light of technological origin from Proxima Centauri revealed no such spectral signatures nor even any candidates. One may determine the laser power that would have been detected from Proxima Centauri, following the approach of Tellis & Marcy (2017). The dominant obstacle against detecting laser light is the flux of light from the star at each wavelength. Proxima Centauri has such low intrinsic luminosity, L = 0.0017 $L_{SUN}$, that laser light of relatively little power can outshine the star as seen by a telescope in the beam.

The detection threshold for laser emission applied to the spectrum was 5 times the photon flux of starlight, shown in Figure 3. The photons gathered in the pixels varied from ~20 photons/pixel at 4000 Å, to 1000 photons/pixel at 5000 Å, to just over 4000 photons/pixel at 6500 Å (Figure 3), representing the approximate number of photons acquired by the 3.6-m telescope in the 900 second exposure. One may translate those photon rates to intrinsic luminosity at each wavelength using the absolute spectrophotometry by Stone (1996) of M dwarf standard stars, shown in Figure 15 of Tellis & Marcy (2017). A laser must be 5x as intense as the stellar luminosity, at a specific wavelength, to meet the detection criterion here. Any such laser emission that spans a range of wavelengths between the width of the instrumental profile (~0.05 Å) and up to 10 Å would be detected.

To determine the required laser power that can be detected, one may adopt a touchstone laser light launcher yielding power requirements that can be scaled to any other laser. The touchstone laser emits a diffraction-limited beam from a laser-launcher with an aperture having a diameter of 10 m. The laser is located in the vicinity of Proxima Centauri. The beam has an opening angle with a first null at an angle, theta = 1.2 $\lambda/D$ from the optical axis, where $\lambda$ is the wavelength of light and $D$ is the laser aperture equal to 10 meters. The opening angle is 0.013 arcsec at 5000 Å. Such a laser beam intercepts the Earth with a circular footprint having area, $A = \pi (1.2 \lambda d/D)^2$, where $d$ is the distance to Proxima Centauri, $d$ = 1.302 pc (Gaia DR3, 2020). At 5000 Å, the footprint at the Earth of the laser beam has a radius of 0.013 au and an area of $1.2 \times 10^{19}$ m$^2$.

For Proxima Centauri, the power required to detect the light from a touchstone laser is 20 kW at 4000 Å, 50 kW at 5000Å, and 120 kW at 6500 Å. These power requirements are remarkably low because of the low luminosity and the proximity of Proxima Centauri. A dim star is easily outshined by a laser, as viewed from within its beam and at its wavelength. This analysis would have detected visible-light lasers of such power if launched from a 10-meter laser launcher at Proxima Centauri that was pointed at Earth. For reference, continuous-wave lasers on Earth can operate for many minutes at power levels over 1 Megawatt.

For assumed laser launchers of smaller aperture, the required laser power would have to be greater to permit detection. For example, a 1-meter laser launcher would have to be 100x more powerful than the requirements above. Short laser pulses from Proxima Centauri would also have been detected. The efficiency of HARPS is 5% and the 3.6-m telescope collecting area is ~10 m$^2$. The detection method here would have detected a pulse of photons at Earth greater than 300 photons/m$^2$ at 4000 Å and 10000 photons/m$^2$ at 5000 Å, for laser pulses of arbitrarily short duration, including sub-nanosecond. The 900 second exposure of the spectrum simply integrates the arriving photons for the duration the pulse. No such laser pulses were detected.

# 8    DISCUSSION AND SUMMARY

Examination of 107 high-resolution spectra of Proxima Centauri, obtained during the years 2004 to 2019, revealed no spectral features of technological origin, notably laser light. Laser beams having power of 20 kW to 120 kW, depending on wavelength, would have been detected, if launched from optics similar to the largest telescopes on Earth of 10-meter diameter. For smaller laser launchers, the power requirements for detection increase inversely as the square of their size. The laser power requirement of 20-120 kW is remarkably small, as solar panels on a roof-top can generate such power, and continuous lasers on Earth can emit such power. The required technology is not exotic, as even modern humanity can construct such lasers.

The Breakthrough Listen team, using the Parkes radio telescope in 2019 April and May, recorded a radio signal at 982 MHz lasting 2.5 hours, calling it "BLC1" (Sheikh et al. 2021, O'Callaghan & Billings 2020, Overbye 2020, Loeb 2021, Koren 2021). The radio signal resides within a narrow frequency range, less than a few Hertz, which is much narrower than emitted by most astrophysical sources such as flares and cyclotron radiation from magnetic fields and stellar coronae. The BLC1 signal had a slight *upward* frequency drift (Sheikh et al. 2021), which is unexpected. The Doppler effect caused by the Earth's orbital motion around the sun would cause a *downward* Doppler frequency shift (i.e. the "eeeeoooooww" pitch of a race car going by – in either direction), thus presenting a puzzle regarding the origin of the frequency drift. The signal appears to come from a region within a cone 20 arcminutes wide, centered on Proxima Centauri, but it could come from one of the sidelobes of the Parkes telescope and receiver.

Among the 107 optical spectra of Proxima Centauri described in this paper, 29 were observed between March and July 2019, fortuitously encompassing the April/May 2019 radio-wave detection of BLC1. None of the 29 optical spectra exhibited any sign of laser light or other technological signals. We thus failed in our effort to support BLC1.

This non-detection of laser light from Proxima Centauri was complicated by interferometric, optical "ghosts" from the HARPS spectrometer. Such undocumented and idiosyncratic behavior of frontier-level optics highlights the challenge of SETI. Every unprecedented "signal" in the data may require weeks or months to discover the quirk of Earth-bound technology (or even nature) that actually produced it. Any fruitless search for the cause of the quirk still leaves conventional explanations viable, albeit unidentified. Most conventional explanations for quirks are located near, or above, the observatory. Thus, *SETI observations would benefit from simultaneous operation of at least two telescopes, separated by at least 1 km, to rule out local false alarms, to provide parallax of signals from satellites, and thus to promote plausibly real signals for further analysis. The second telescope can be smaller, sufficient to produce 3-sigma confirmation of 10-sigma detections.*

Proxima and Alpha Centauri are optimal targets for continued SETI observations. Any technological entities there could use a radio telescope or an interferometric infrared telescope (e.g. Dandumont et al. 2020) to detect signals coming from Earth, such as our radar beacons or city lights (Zuckerman 2002). Detecting such signals from Earth starting 80 years ago, they could deploy a probe for closer surveillance (Hippke 2021, Gertz 2021). Travelling at a mere 1/20 the speed of light, a probe could traverse the 4.2 light years, and be near the Solar System already.

The ensuing communication between the probe here and its home-base at Proxima or Alpha Centauri may be accomplished with lasers emitting infrared, optical, or UV frequencies. Tight laser beams offer private communication, 10 GHz bandwidth, and vital defensive stealth. A modest, meter-size laser at Proxima Centauri could spotlight a region smaller than the Earth-Sun distance near our Solar System, enabling their transmission to avoid Earth entirely. This scenario is consistent with the non-detection of laser light found here.

A variation of this scenario involves using the Sun as a gravitational lens to amplify the laser communication (Hippke 2021). We have carried out observations of the Solar gravitational lens point of both Proxima and Alpha Centauri to search for ongoing laser communication between the two neighboring star systems (Marcy & Hippke 2021).

## 9    ACKNOWLEDGMENTS

DATA AVAILABILITY

This paper is based on data products, reduced spectra and raw images, obtained with ESO Telescopes at the La Silla Paranal Observatory. All data are available to the public at the ESO archive:
https://www.eso.org/sci/facilities/lasilla/instruments/harps/tools/archive.html
The data can be retrieved by entering "Proxima Centauri" into the archive website.

The spectra used in this paper were obtained on the ESO 3.6-meter telescope and the HARPS spectrometer between 2004 and 2019 within four programs. The first program was under the direction of PI/Co-I Mayor, Benz, Bertaux, Bouchy, Pepe, Perrier, Queloz, Sivan, Udry, Delfosse, Forveille, Santos, Moutou, Barge, Deleuil, Schmidt, Guillot, Lovis, and Bonfils, titled "A high-precision systematic search for exoplanets in the Southern Sky." The second program had PI/Co-I Bonfils, Delfosse, Forveille, Gillon, Perrier, Santos, and Udry titled, "Transits of Telluric Exoplanets Orbiting M Stars." The third and fourth programs were led by PI Anglada-Escude and others to study M dwarfs, including Proxima Centauri.

The spectra were obtained under four programs. The first was ID 072.C-0488E, HARPS-Guaranteed Time, led by Michel Mayor, contributing 19 spectra obtained between May 2005 and July 2008, with integration times between 450 and 900 s. The second and third programs were ID 082.C-0718(B) and ID183.C-0437(A), both led by X.Bonfils that contributed 8 and 46 spectra with exposure times of 15 minutes. The fourth program, 191.C-0505(A), was led by G. Anglada-Escuda contributing 70 spectra between May 4 and May 16, 2013, and another 23 spectra between Dec 30 2013 and Jan 10 2014. We are grateful for the publicly archived high-resolution spectra obtained with the HARPS spectrometer maintained by the European Southern Observatory (ESO). We thank the PI and CoI's of the HARPS program for the design, construction, observations, and reduction of the HARPS spectra, namely, Drs. Mayor, Benz, Bertaux/Bouchy, Pepe, Perrier, Queloz, Sivan, Udry, Delfosse, Forveille, Santos, Moutou, Barge, Deleuil, Schmidt, Guillot, Lovis, Bonfils, Delfosse, Forveille, Gillon, Perrier, Santos, and Anglada-Escude et al.

This paper has been typeset from Microsoft WORD document prepared by the author.